# Structure of skimmed milk during ultrafiltration process probed by in-situ small angle X-ray scattering

Pignon[1] F., Belina[1,2] G., Narayanan[2] T., Paubel[1] X., Magnin[1] A., Gésan-Guiziou[3] G.

[1]Laboratoire de Rhéologie, Université Joseph Fourier Grenoble I, Institut National Polytechnique de Grenoble, CNRS, UMR 5520, BP 53, F-38041, Grenoble, Cedex 9 France,
[2]European Synchrotron Radiation Facility, BP 220, F-38043, Grenoble Cedex 9, France,
[3]UMR 1253 INRA Agrocampus, Science et Technologie du Lait et de l'Œuf, 65 rue de Saint Brieuc, F-35042, Rennes Cedex, France.

pignon@ujf-grenoble.fr

**Abstract**
The stability and mechanism underlying the formation of deposits of casein micelles during ultrafiltration process were investigated by small-angle and ultra small-angle X-ray scattering (SAXS and USAXS). The casein micelle dispersions consisted of phospho-caseinate model powders and fresh skimmed milk and the measurements probed length scales ranging from 1 to 2000 nm. Rheometric and frontal filtration measurements were combined with SAXS to establish the relationship between the rheological behavior of deposits (shear and/or compression) and the corresponding microstructure. The results obtained clearly demonstrate that the equilibrium structure of casein micelles is globular with a radius of gyration ($R_g \sim 100$ nm). The internal structure of the micelles is more likely an unfolded state of the constituent proteins rather than globular sub-micelles (Fig. 2). In-situ ultrafiltration experiments under SAXS relate the reduction in the filtration rate to the evolution of microstructure near the membrane. The specific resistance shows a quick linear increase before slowing down. Correspondingly, the scattering pattern in the close vicinity of the membrane becomes anisotropic for long times filtration (Fig. 1) indicating the deformation of the globular micelles. From absolute SAXS intensities, the concentration of the micelles in the deposit can be deduced reliably (Fig. 3).

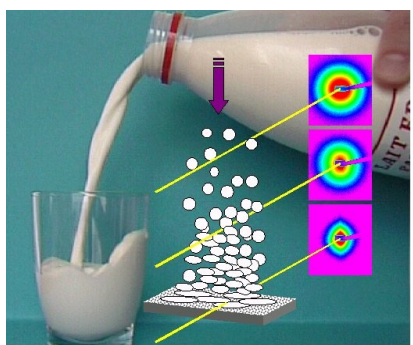

Fig. 1: In-situ SAXS of deposits probed during ultrafiltration process of casein micelles dispersions.

### Introduction

In the dairy industry, the ultrafiltration process is widely used for the fractionation of skimmed milk. One of the main limitations of the membrane separation performance is the accumulation of matter at the membrane surface. Previous work (Gésan-Guiziou *et al.*, 1999) concerning the microfiltration of skimmed milk showed that the transition from stable filtration flux to unstable condition is mainly governed by the formation of a deposit on the membrane. Consequently, it is of capital interest to understand the mechanisms of deposit formation, its structural arrangement and its rheological behavior in order to be able to predict its apparition, to control and to reduce its effects (Pignon *et al.* 2004). Furthermore, in order to assess the filtration performance and to elucidate the limits of the stability zone, the structural properties of the deposits must be combined with permeation measurements and the determination of reversibility condition. In the case of skimmed milk, an important aspect need to be considered is the structure of casein micelles itself, which are primarily aggregates consisting of caseins (main proteins of milk) and minerals.

## Results and Conclusions

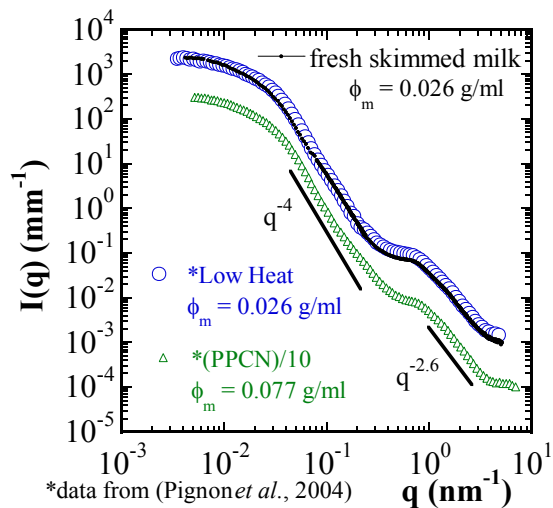

Fig. 2: Static structure of casein micelles dispersions investigated by SAXS and USAXS at T = 25 °C and pH = 6.6.

*data from (Pignon *et al.*, 2004)

*Static structure of casein micelles*

The casein micelles dispersions (mass concentration of casein micelles ($\phi_m$)) were prepared from fresh skimmed milk, purified condensed native phosphocaseinate micelles containing 84.5% of native micellar casein (PPCN) and weakly denatured skimmed milk powder (Low Heat) containing soluble proteins and mineral salts in addition to 28.5% of micellar casein. Adequate suspending phases have been used to obtain the static structure of the native casein micelles investigated by SAXS and USAXS measurements (Fig. 2). The scattering Intensity $I(q)$ over the low $q$ branch (corresponding to the larger structure) can be fitted by the Debye-Büche function, and decays by a $q^{-4}$ power law. These clearly demonstrate that the larger units consist of globular objects micelles with a radius of gyration ($R_g \sim 100$ nm). The high $q$ region shows a significantly different behavior with $I(q)$ decreasing like $q^{-2.6}$. This lower structural level ($R_g \sim 5$ nm) is primarily due to the nanometric calcium phosphate particles reticulated in the protein matrix within in the large globular micelles. The high $q$ region of the scattering intensity behavior suit well with a model which describes casein micelles as a relatively uniform matrix containing a disordered micelle calcium phosphate (Holt *et al.*, 2003).

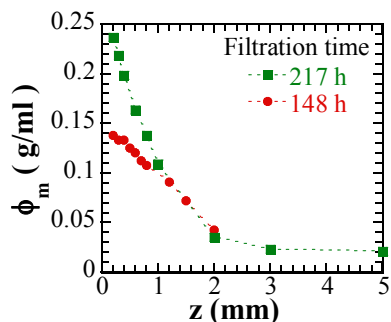

Fig. 3: Concentration profiles in deposits formed by frontal filtration of Low Heat ($\phi_m$ = 0.081 g/ml, pH = 6.6, transmembrane pressure = $10^5$ Pa, and T = 25 °C).

*In situ SAXS during ultrafiltration process*

Frontal filtration cells were specially developed to probe the microstructure of deposits by SAXS during the separation process with an accuracy of 0.1 mm. In-situ filtration experiments relate the reduction in the filtration rate to the evolution of microstructure near the membrane. The 2d SAXS patterns (Fig. 1) are noticeably anisotropic very close to the membrane (z = 0.2 mm) indicating the deformation of the globular micelles by the applied filtration pressure. from the absolute SAXS intensities, the concentration profile (Fig. 3) and the anisotropy as a function of the distance z to the membrane can be deduced reliably (Pignon *et al.*, 2004). These results suggests that the mechanisms responsible for the reduction of permeation flux is a cumulative effect of a high increase of concentration and of the deformation of the micelles in the immediate vicinity of the membrane (Belina, 2005).